\documentclass{sf2a-conf}
\usepackage{graphicx}

\begin{document}
\TitreGlobal{SF2A 2007}
\title{Magnetism and rotation in Herbig Ae/Be stars}


\author{E. Alecian} \address{Royal Military College of Canada, Kingston, Ontario, Canada}
\author{G.A. Wade$^1$} 
\author{C. Catala} \address{Observatoire de Paris, LESIA, Meudon, France}
\author{C. Folsom$^1$}
\author{J. Grunhut$^1$}
\author{J.-F. Donati} \address{Laboratoire d'Astrophysique, Observatoire Midi-Pyr\'en\'ees, France}
\author{P. Petit$^3$}
\author{S. Bagnulo} \address{Armagh Observatory, Northern Irland}
\author{T. Boehm$^3$}
\author{J.-C. Bouret} \address{Laboratoire d'Astrophysique de Marseille, France}
\author{J.D. Landstreet} \address{Dept. of Physics \& Astronomy, University of Western Ontario, Canada}

\runningtitle{Magnetism and rotation in Herbig Ae/Be stars}

\setcounter{page}{1}

\index{E. Alecian}
\index{G.A. Wade} 
\index{C. Catala}

\maketitle

\begin{abstract}
Among the main sequence intermediate mass A and B stars, around 5\% host large-scale organized magnetic fields. Most of these stars are very slow rotators compared to their non-magnetic counterparts, and show photospheric abundance anomalies. They are referred to as the Ap/Bp stars. One of the greatest challenges, today is to understand the origin of their magnetic field and their slow rotation. The favoured hypothesis is a fossil origin of the magnetic field, in which the magnetic fields of Ap/Bp stars are relics of those which existed in the parental molecular clouds during the formation. This implies that the magnetic field must survive all the initial phases of the stellar evolution and especially the pre-main sequence (PMS) phase. This is consistent with the general belief that magnetic braking occurs during the PMS phase, which sheds angular momentum and slows the rotation of these stars. In this context, we proceeded with a survey of a sample of around 50 PMS Herbig Ae/Be stars, using the new spectropolarimeter ESPaDOnS at the CFHT, in order to study the magnetic field and the rotation velocity of these stars. This talk reviews the results of our survey, as well as their consequences for the origin of the magnetic fields and the evolution of the rotation of intermediate mass stars during the PMS phase.
\end{abstract}

%

\section{Introduction}

\subsection{Magnetism and rotation in the main sequence A and B stars}

Between about 1.5 and 10 M$_{\odot}$, at spectral types A and B, about 5 \% of main sequence (MS) stars have magnetic fields with characteristic strengths of about 1kG. Such stars also show important chemical peculiarities and are thus usually called the magnetic chemically peculiar Ap/Bp stars. The strength of the magnetic fields of these stars cannot be explained by an envelope dynamo as in the sun. Until now, the most reliable hypothesis has been to assume a fossil origin for these magnetic fields. This hypothesis implies that the stellar magnetic fields are relics from the field present in the parental interstellar cloud. It also implies that magnetic fields can (at least partially) survive the violent phenomena accompanying the birth of stars, and can also remain throughout their evolution and until at least the end of the MS, without regeneration.

According to the fossil field model, magnetic fields should reside in some pre-main sequence (PMS) stars of intermediate mass, the so-called Herbig Ae/Be stars. However no magnetic field was observed up to recently in these stars (except HD~104237, Donati et al. 1997). Can we obtain some observational evidence of the presence of magnetic fields during the PMS phase of evolution, as predicted by the fossil field hypothesis? If some Herbig Ae/Be stars are discovered to have magnetic fields, is the fraction of magnetic to non-magnetic Herbig stars the same as the fraction for main sequence stars? Is the magnetic field in Herbig stars strong enough to explain the strength of that of Ap/Bp stars?

Chemical peculiarities and magnetism are not the only characteristic properties observed in the Ap/Bp stars. Most magnetic MS stars have rotation periods (typically of a few days) that are several times longer than the rotation periods of non-magnetic MS stars (a few hours to one day). It is usually believed that magnetic braking, in particular during PMS evolution, when the star can exchange angular momentum with its massive accretion disk, is responsible for this low rotation (St{\c e}pie{\'n} 2000, St{\c e}pie{\'n} \& Landstreet 2002). An alternative involves a rapid dissipation of the magnetic field during the early stages of PMS evolution for the fastest rotators, due to strong turbulence induced by rotational shear developed under the surface of the stars, as the convection do in the solar-type stars (see e.g. Ligni\`eres et al. 1996). In this scenario, only slow rotators could retain their initial magnetic fields, and evolve as magnetic stars to the main sequence. So the question to be addressed is the following: does the magnetic field control the rotation of the star, or else does the rotation of the star control the magnetic field? We propose that this question can be answered by studying rotation and magnetic fields in Herbig Ae/Be stars.

\subsection{The Herbig Ae/Be stars}

The Herbig Ae/Be stars are intermediate-mass pre-main sequence stars, and therefore the evolutionary progenitors of the MS A and B stars. They are distinguished from the classical Be stars by their IR emission and the association with nebulae, characteristics which are due to their young age. 

They display many observational phenomena often associated with magnetic activity. First, high ionised lines are observed in the spectra of some stars (e.g. Bouret et al. 1997, Roberge et al. 1001), and X-ray emission have been detected, coming from some Herbig stars (e.g. Hamaguchi et al. 2005). In active cool stars, many of these phenomena are produced in hot chromospheres or coronae. Some authors mentioned rotational modulation of resonance lines which they speculate may be due to rotation modulation of winds structured by magnetic field (Praderie et al. 1986, Catala et al. 1989, Catala et al. 1991, Catala et al. 1999). 

In the literature we find many clues of the presence of circumstellar disks around these stars, from spectroscopic data showing  strong emission, and also from photometric data (e.g. Mannings \& Sargent 1997, Mannings \& Sargent 2000). Recently, using coronagraphic data and interferometric data, some authors have also found direct evidence of circumstellar disks around these stars (Grady et al. 1999, Grady et al. 2000, Eisner et al. 2003). A careful study of these disks shows that they have similar properties to the disk of their low mass counterpart (Natta et al. 2001), the T Tauri stars, whose the emission lines are explained by magnetospheric accretion models (Koenigl 1991, Muzerolle et al. 1998, Muzerolle et al. 2001). Finally Muzerolle et al. (2004) have sucessfully applied their magnetospheric accretion model to Herbig stars to explain the emission lines in their spectra. 

For all these reasons we suspect that the Herbig stars may host large-scale magnetic fields that should be detectable with current instrumentation. However, many authors tried to detect such fields without much success (Catala et al. 1993, Catala et al. 1999, Donati et al. 1997, Hubrig et al. 2004, Wade et al. 2007). But in 2005, a new high-resolution spectropolarimeter, ESPaDOnS, has been installed at the canada-france-hawaii telescope (CFHT). We therefore decided to proceed to survey many Herbig stars in order to investigate rotation and magnetism in the pre-main sequence stars of intermediate mass.

\section{Observations and reduction}

\subsection{Our sample}

\begin{figure}
\centering
\includegraphics[height=8cm]{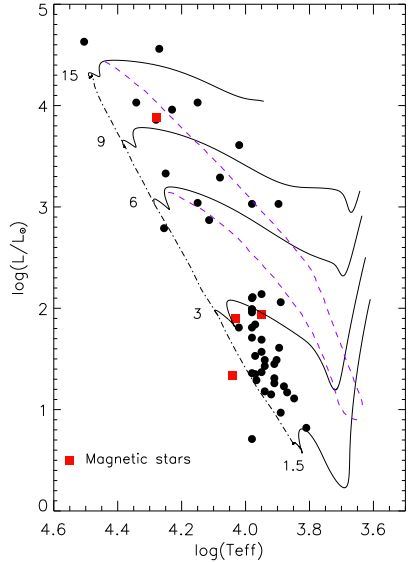}
\caption{Herbig stars plotted in a HR diagram. The red squares are the magnetic Herbig stars. The PMS evolutionary tracks are plotted for different masses (full lines). The birthline for 10$^{-5}$ and 10$^{-4}$ $M_{\odot}$.yr$^{-1}$ mass accretion rate are plotted in dashed line (Palla \& Stahler 1993). The dash-dotted line is the ZAMS.}
\label{fig:hr}
\end{figure}

We have selected Herbig stars in the catalogues from Th\'e  et al. (1994) and Vieira et al. (2003) with a visual magnitude brighter than 12. Our sample contain 55 stars which have masses ranging from 1.5 M$_{\odot}$ to 15 M$_{\odot}$ with all ages between the birthline and the zero-age main sequence (ZAMS). In Fig. \ref{fig:hr} are plotted the stars of our sample in an HR diagram, as well as the evolutionary tracks computed with the CESAM code (Morel 1997), and the birthlines computed by Palla \& Sthaler (1993) with two mass accretion rates during the protostellar phase : 10$^{-5}$~M$_{\odot}$.yr$^{-1}$ and 10$^{-4}$~M$_{\odot}$.yr$^{-1}$.

\subsection{Observations and reduction}

Our data were obtained using the high resolution spectropolarimeter ESPaDOnS installed on the 3.6 m Canada-France-Hawaii Telescope (Donati et al. 2007, in preparation) during many scientific runs.

We used this instrument in polarimetric mode, generating spectra of 65000 resolution. Each exposure was divided in 4 sub-exposures of equal time in order to compute the optimal extraction of the polarisation spectra (Donati et al. 1997, Donati et al. 2007 in prep.). We recorded only circular polarisation, as the Zeeman signature expected in linear polarisation is about one order of magnitude lower than circular polarisation. The data were reduced using the "Libre ESpRIT" package especially developed for ESPaDOnS, and installed at the CFHT (Donati et al. 1997, Donati et al. 2007, in preparation). After reduction, we obtained the intensity Stokes $I$ and the circular polarisation Stokes $V$ spectra of the stars observed.

We then applied the Least Squares Deconvolution procedure to all spectra (Donati et al. 1997), in order to increase our signal to noise ratio. This method assumes that all lines of the intensity spectrum have a profile of similar shape. Hence, this supposes that all lines are broadened in the same way. We can therefore consider that the observed spectrum is a convolution between a profile (which is the same for all lines) and a mask including all lines of the spectrum. We therefore apply a deconvolution to the observed spectrum using the pre-computed mask, in order to obtain the average photospheric profiles of Stokes $I$ and $V$. In this procedure, each line is weighted by its signal to noise ratio, its depth in the unbroadened model and its Land\'e factor. For each star we used a mask computed using "extract stellar" line lists obtained from the Vienna Atomic Line Database (VALD)\footnote{http://www.astro.univie.ac.at/$\sim$vald/}, with effective temperatures and $\log g$ suitable for each star (Wade, Alecian et al. 2007, in prep.). We excluded from this mask hydrogen Balmer lines, strong resonance lines, lines whose Land\'e factor is unknown and emission lines. The results of this procedure are the mean Stokes $I$ and Stokes $V$ LSD profiles (Fig. \ref{fig:compil}).

\section{Results}

\subsection{Discovery of magnetic fields in Herbig stars}

\begin{figure}
\centering
\includegraphics[width=6.8cm, angle=90]{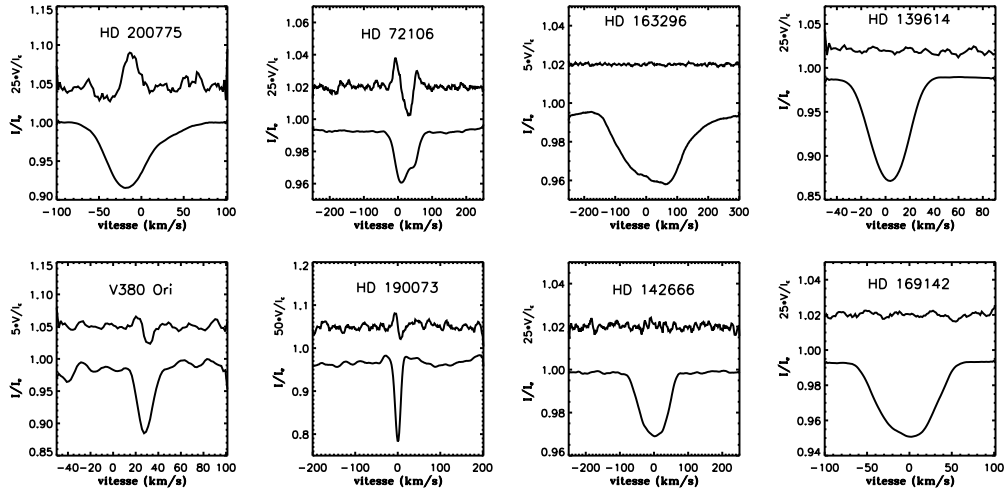}
\caption{Stokes $I$ (bottom) and Stokes $V$ (up) LSD profiles plotted for the 4 magnetic stars (right and middle) and two undetected stars (left). Note the amplification factor in $V$.}
\label{fig:compil}
\end{figure}

Thanks to the high performance of the instrument ESPaDOnS and to the LSD method, {\bf we have discovered four new magnetic Herbig Ae/Be stars} (Wade et al. 2005, Catala et al. 2006, Alecian et al. 2007). In Fig. \ref{fig:compil} is plotted the Stokes $I$ and $V$ profiles of each new discovered magnetic Herbig Ae/Be star : HD~200775, HD~72106, V380~Ori and HD~190073, and of four stars in which a magnetic field has not been detected (the undetected stars, hereafter). Contrary to the undetected stars, the Stokes $V$ profiles of the magnetic stars are not null and display a strong Zeeman signature, of the same width of the photospheric $I$ profile, characteristic of the presence of a magnetic field in the stars.

These 4 detections among our sample of 55 stars lead to the conclusion that {\bf 7\% of Herbig Ae/Be stars are magnetic}. The projection of the distribution of magnetic and non-magnetic main sequence A and B stars on the pre-main sequence phase, assuming a fossil field hypothesis, predict that between 5 and 10 \% of Herbig stars should be magnetic, which is consistent with our observations. We therefore bring a new strong argument in favoured of the fossil field hypothesis (Wade, Alecian et al. 2007, in prep.).

\subsection{Topology and intensity of the magnetic fields}

We determined the topology and the intensity of the magnetic fields of the four magnetic Herbig stars in order to compare them to the magnetic MS A and B stars. With this aim, we used the oblique rotator model described by Stift (1975). We consider a dipole placed at a distance $d_{\rm dip}$ on the magnetic axis of a spherical rotating star with a magnetic intensity at the pole $B_{\rm P}$. The rotation axis of the star is inclined at an angle $i$ with respect to the line of sight and makes an angle $\beta$ with the magnetic axis.

\begin{table}
\caption{Fundamental, geometrical and magnetic parameters of the magnetic Herbig Ae/Be stars. References : 1 : Alecian et al. (2007), 2 : Folsom et al. (2007), in prep., 3 : Alecian et al. (2007), in prep., 4 : Catala et al. (2006).}
\label{tab:fp}
\centering
\begin{minipage}[t]{\linewidth}
\begin{tabular}{lcccccccccc}
\hline
\hline
Star   & S.T. & $v\sin i$                & age     & P    & $B_{\rm P}$ & $\beta$ ($^{\circ}$) & $i$ ($^{\circ}$) & $d_{\rm dip}$ & $B_{\rm P (ZAMS)}$ & Ref. \\
         &         & (km.s$^{\rm -1}$) & (Myr)   & (d)  & (kG)             &                                 &                         & $R_*$             & (kG)                          &        \\
\hline
HD 200775 & B2 & 26  & 0.1 & 4.328         & 1             & 78            & 13            & 0.1 & 3.6                                             & 1 \\
HD 72106   & A0 & 41  & 10  & 0.63995     & 1.5          & 58            & 23            & 0    & 1.5                                             & 2 \\
V380 ori\footnote{Work in progress. We need more data of V380~Ori to choose between the 7.6 and the 9.8 periods. Therefore two solutions are possible for $\beta$ and $i$}      & A2 & 9.8 & 2.8 & $[7.6,9.8]$ & 1.4          & $[90,85]$ & $[36,49]$ & 0    & 2.4 & 3 \\
HD 190073\footnote{Although we have observed HD 190073 over more that 2 years, no variation of the Stokes $V$ profile has yet been detected.} & A2 & 8.5 & 1.5 &                   & $[0.1,1]$ & $[0,90]$   & $[0,90]$   &       & $[0.3,3]$                                    & 4 \\
\hline
\end{tabular}
\end{minipage}
\end{table}

According to Landi degl'Innocenti \& Landi degl'Innocenti (1973), in the weak field approximation, the Stokes $V$ profile is proportional to the magnetic field projected onto the line of sight and integrated over the surface of the star ($B_{\ell}$, the longitudinal magnetic field, hereafter). As the star rotates, the visible magnetic changes, resulting in variation of $B{_\ell}$. Therefore the Stokes $V$ profile changes with the rotation phase.

\begin{figure}
\centering
\includegraphics[width=6.8cm, angle=90]{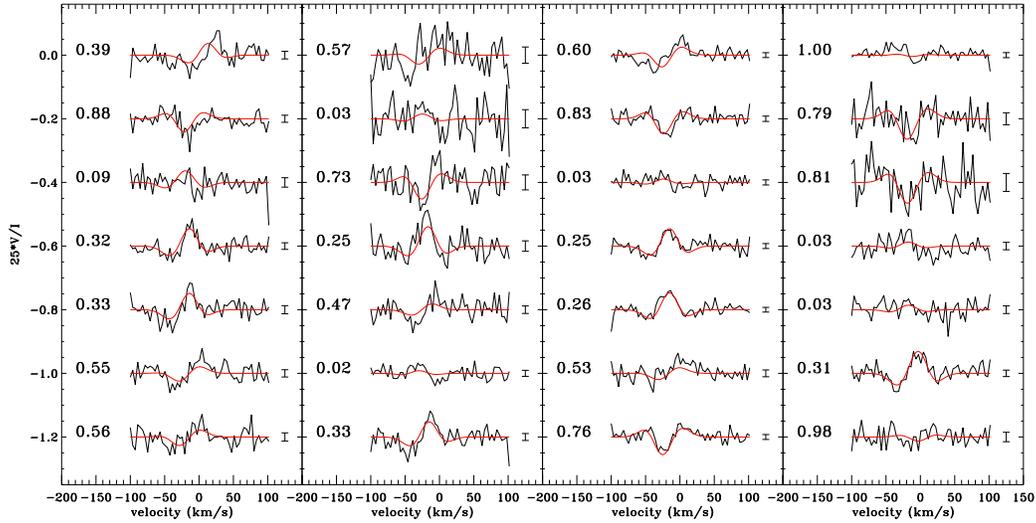}
\caption{Stokes $V$ profiles of HD~200775 superimposed to the synthetic ones, corresponding to our best fit. The rotation phase and the error bars are indicated on the left and on the right of eache profiles, respectively (Alecian et al. 2007).}
\label{fig:fitallv}
\end{figure}

In order to determine the geometrical and magnetic parameters $i$, $\beta$, $B_{\rm P}$ and $d_{\rm dip}$, as well as the rotation period $P$ of the star, we observed the stars at different rotation phases and fit simultaneously all the Stokes $V$ profiles observed for each star. With this aim we calculated a grid of $V$ profiles, using the oblique rotator model, for each date of observations, varying the five parameters. Then, for each star, we applied a $\chi^2$ minimisation to find the best model which matches simultaneously all the $V$ profiles observed. Fig. \ref{fig:fitallv} shows the result of our fitting procedure for one star : HD 200775. The synthetic Stokes $V$ are superimposed on the observed ones (Alecian et al. 2007).

The values of the geometrical and magnetic parameters are summarized in Table \ref{tab:fp} for each stars. In the case of HD 190073, the topology and the intensity of its magnetic field are not constrained, because, during 2 years observations, the Stokes $V$ profile has not been observed to vary. There are 3 possible explanations for this : the inclination $i$ is very small, the obliquity angle $\beta$ is very small, or the rotation period of the star is very long. More observations will allow us to discard two of these solutions. However the stability of the magnetic field over more than 2 years and the shape of the Stokes $V$ profiles lead us to the conclusion that this star hosts a large-scale fossil magnetic field (Catala et al. 2006).

The success of our fitting procedure for the 3 stars HD 200775, HD 72106 and V380 ori, as well as our discussion on HD 190073, lead to the conclusion that the {\bf magnetic Herbig Ae/Be stars host globally dipolar magnetic fields}, similar to the Ap/Bp stars.

Assuming the conservation of magnetic flux during the PMS evolution, and using the radius of the stars and their predicted radius on the ZAMS for the same mass, we can estimate the magnetic intensity at their surface they will have when they will reach the ZAMS (see Table \ref{tab:fp}). We found intensities ranging from 300~G to 3.6 kG, which is very close to what is observed in the Ap/Bp stars. Hence we bring 2 more new strong arguments in favour of the fossil field hypothesis.

\section{Conclusions}

We used the new spectropolarimeter ESPaDOnS installed at the CFHT to proceed in a survey of the Herbig stars, in order to investigate their rotation and magnetic field. We discovered four magnetic stars whose field topology is similar to the MS magnetic A-B stars. We also show that the magnetic intensities of these fields and the proportion of magnetic Herbig stars can explain the magnetic intensity and the proportion of magnetic fields among the MS stars, in the context of fossil field model. We therefore bring fundamental arguments in favour of this hypothesis.

The four magnetic Herbig stars are slow rotators ($v\sin i < 41$ km.s$^{-1}$) which supports that magnetic Herbig Ae/Be stars are the progenitors of the magnetic Ap/Bp stars. Among these magnetic stars two have very low $v\sin i$ ($< 10$ km.s$^{-1}$) and are very young (age$ < 2.8$ Myr). Assuming these stars are true slow rotators, this implies that there exists a braking mechanism which acts very early during the PMS evolution of the intermediate mass stars. We could think that this braking mechanism has a magnetic origin, although among the undetected stars we also observe same stars with small $v\sin i$ ($\sim 15$ km.s$^{-1}$). The nature of the braking mechanism requires addition study.

\end{document}